\documentclass[twocolumn,aps,prl,superscriptaddress,floatfix,preprintnumbers,showpacs]{revtex4-1}
\usepackage{amsmath, amssymb}
\usepackage{bm}
\usepackage{graphicx}
\usepackage{subfigure}
\usepackage[dvipdfmx,colorlinks=true,linkcolor=blue,citecolor=blue,urlcolor=black]{hyperref}

\begin{document}
\title{Impact of Rattlers on Thermal Conductivity of a Thermoelectric Clathrate: A First-Principles Study}

\author{Terumasa Tadano}
\email{tadano@cms.phys.s.u-tokyo.ac.jp}
\affiliation{Department of Physics, The University of Tokyo, Tokyo 113-0033, Japan}

\author{Yoshihiro Gohda}
\affiliation{Department of Materials Science and Engineering, Tokyo Institute of Technology, Yokohama 226-8502, Japan} 

\author{Shinji Tsuneyuki}
\affiliation{Department of Physics, The University of Tokyo, Tokyo 113-0033, Japan}
\affiliation{Institute for Solid State Physics, The University of Tokyo, Kashiwa 277-8581, Japan}

\begin{abstract}    
We investigate the role of rattling guest atoms on the lattice thermal-conductivity of a type-I clathrate Ba$_{8}$Ga$_{16}$Ge$_{30}$ by first-principles lattice dynamics.
Comparing phonon properties of filled and empty clathrates, we show that rattlers cause 10-fold
reductions in the relaxation time of phonons by increasing the phonon-phonon scattering probability.
Contrary to the resonant scattering scenario, the reduction in the relaxation time occurs in a wide
frequency range, which is crucial for explaining unusually low thermal-conductivities of clathrates.
We also find that the impact of rattlers on the group velocity of phonons is secondary because the
flattening of phonon dispersion occurs only in a limited phase space in the Brillouin zone.
\end{abstract} 
\maketitle


An intermetallic clathrate is an inclusion complex where guest atoms are 
enclosed in cavities formed by the host crystal lattice~\cite{rogl}.
One of the most characteristic features of clathrates is an unusually low 
lattice thermal-conductivity $\kappa_{\mathrm{L}}$ ($\sim 1$ W/mK)~\cite{Takabatake2014},
which makes clathrates promising for thermoelectric applications
with high figure of merit $ZT = \sigma S^{2}T/(\kappa_{\mathrm{c}}+\kappa_{\mathrm{L}})$. 
Here, $T$ is the absolute temperature, $\sigma$ is the electrical conductivity, $S$ is the Seebeck coefficient, 
and $\kappa_{\mathrm{c}(\mathrm{L})}$ is the thermal conductivity by electrons (phonons), respectively.
Since $\kappa_{\mathrm{L}}$ of a clathrate is intrinsically low without introducing 
micro/nanostructures such as grain boundaries or nanoscale precipitates, 
semiconductor clathrates are one of prototype materials which follow the phonon glass/electron crystal (PGEC) concept proposed by Slack~\cite{Slack}. 


The origin of low $\kappa_{\mathrm{L}}$ of host-guest structures such as clathrates and skutterudites 
has commonly been attributed to the ``rattlers'', i.e. guest atoms loosely bound inside oversized cages~\cite{Cohn1999,Tse2001,Nolas1998,Toberer2011}. 
However, its actual role is not fully understood.
In the simple kinetic theory, the lattice thermal conductivity is given by
\begin{equation}
\kappa_{\mathrm{L}} = \frac{1}{3}Cv^{2}\tau,
\end{equation}
where $C$ is the lattice specific heat, $v$ is the average group velocity, 
and $\tau$ is the average relaxation time of phonons, respectively.
Historically, the reduction in $\kappa_{\mathrm{L}}$ has been attributed to resonant scatterings by localized rattling modes~\cite{Cohn1999}, which reduce $\tau$ of 
heat-carrying acoustic modes in a limited energy region near avoided-crossing points.
This mechanism was originally introduced to explain $\kappa_{\mathrm{L}}$ observed in a solid solution of 
KCl and KNO$_{2}$~\cite{Pohl1962}, and it was also applied to a clathrate hydrate assuming the guest as an isolated point defect~\cite{Tse2001}.
However, although the resonant phonon scattering has repeatedly been employed to explain low $\kappa_{\mathrm{L}}$'s of host-guest structures~\cite{Cohn1999,Tse2001,Nolas1998}, 
little attention has been paid to the validity of that mechanism itself.
Recently, the validity of resonant scattering was questioned by experimental and theoretical studies in skutterudites \cite{Koza2008,Li2014}.


Another important role of rattlers which has recently been recognized is its impact on the group velocity~\cite{Christensen2008,Christensen2010}.
On the basis of inelastic neutron scattering (INS) of a type-I clathrate Ba$_{8}$Ga$_{16}$Ge$_{30}$ (BGG), 
Christensen \textit{et al.}~\cite{Christensen2008} claimed that a major effect of rattlers is to reduce $v$ of acoustic phonons at the avoided-crossing points rather than to reduce $\tau$. 
This mechanism clearly conflicts with the resonant scattering scenario, where $\tau$ is the main source of low $\kappa_{\mathrm{L}}$.
It is still an open question that which of the group velocity $v$ or the relaxation time $\tau$ is mainly affected by rattlers, 
which should be understood precisely for further reducing $\kappa_{\mathrm{L}}$, and hereby enhancing $ZT$.


In this Letter, we analyze the effect of rattlers on $\kappa_{\mathrm{L}}$ in a type-I clathrate BGG from first principles
to answer the questions raised above. 
In particular, we extract harmonic and anharmonic interatomic force constants (IFCs) from \textit{ab initio} calculations based on density-functional theory (DFT), and then we calculate the phonon frequency and phonon lifetime of filled and empty clathrates. 
As will be shown in the following, we observe a 10-fold reduction in $\tau$ and hereby in $\kappa_{\mathrm{L}}$ due to rattlers, whereas the change in the group velocity is less significant. 
We also show that the reduction in $\tau$ occurs in a wide frequency range, thus indicating that the phonon scatterings in BGG is nonresonant.


For a reliable estimation of phonon relaxation times, one needs to consider dominant phonon scattering processes.
Generally, phonons can be scattered by other phonons, charged carriers, disorders, isotopes, and grain boundaries~\cite{kaviany_book}. 
In clathrates with off-center rattlers, phonon scatterings due to tunneling between the two-level system should also be considered
to explain glasslike $\kappa_{\mathrm{L}}$ at low temperature~\cite{Takabatake2014}.
In this study, we only consider the three-phonon interactions which are dominant in a relatively high-temperature range for on-center systems which we are interested in. Then, by considering the lowest-order perturbation of cubic anharmonicities, which has successfully been applied for many solids~\cite{Tian2012,Lindsay2013}, the linewidth of phonon mode $q$ is given as~\cite{Maradudin1962}
\begin{align}
\Gamma_{q}(\omega) &= \frac{\pi}{2N}\sum_{q',q''} | V_{3}(-q,q',q'')|^{2} \notag \\
&\hspace{5mm} \times \left[ (n_{q'} + n_{q''} + 1)\delta(\omega - \omega_{q'} - \omega_{q''}) \notag  \right. \\
&\hspace{11mm} \left. - 2(n_{q'} - n_{q''}) \delta(\omega - \omega_{q'} + \omega_{q''}) \right] .
\label{eq:Gamma}
\end{align}
Here, $\omega_{q}$ is the phonon frequency, $N$ is the number of $\bm{q}$ points,  
$n_{q} = 1 / (e^{\beta\hbar\omega_{q}}-1)$ is the Bose-Einstein distribution function, $\beta=1/kT$ with the Boltzmann constant $k$, and $\hbar$ is the reduced Planck constant, respectively.
In Eq.~(\ref{eq:Gamma}) and in the following, we use the variable $q$ defined by $q = (\bm{q},j)$ and $-q = (-\bm{q}, j)$ where $\bm{q}$ is the crystal momentum and $j$ is the branch index of phonons. 
The matrix element for three phonon interaction $V_{3}$ is given as
\begin{align}
V_{3}(q,q',q'') &= \frac{1}{N}\sqrt{\frac{\hbar^{3}}{8\omega_{q}\omega_{q'}\omega_{q''}}} \notag \\ 
&\hspace{5mm} \times \sum_{\{\ell,\alpha,\mu\}} \Phi_{\mu_{1}\mu_{2}\mu_{3}}(\ell_{1}\alpha_{1};\ell_{2}\alpha_{2};\ell_{3}\alpha_{3})  \notag \\
&\hspace{5mm} \times \frac{e_{\alpha_{1}}^{\mu_{1}}(q)e_{\alpha_{2}}^{\mu_{2}}(q')e_{\alpha_{3}}^{\mu_{3}}(q'')}{\sqrt{M_{\alpha_{1}}M_{\alpha_{2}}M_{\alpha_{3}}}} \notag \\ 
&\hspace{5mm} \times e^{i(\bm{q}\cdot\bm{r}(\ell_{1})+\bm{q}'\cdot\bm{r}(\ell_{2})+\bm{q}''\cdot\bm{r}(\ell_{3}))},
\end{align}
where $e_{\alpha}^{\mu}(q)$ is the polarization vector of atom $\alpha$ along $\mu$ direction which corresponds to phonon mode $q$, $M_{\alpha}$ is the atomic mass of atom $\alpha$, and $\bm{r}(\ell)$ is the coordinate of $\ell$th cell. 
The phonon frequency $\omega_{q}$ and the polarization vector $\bm{e}(q)$ can be obtained straightforwardly by diagonalizing the dynamical matrix.
Once the phonon linewidths are obtained, the lattice thermal conductivity is estimated by the Boltzmann transport equation (BTE)
within the relaxation time approximation (RTA)
\begin{equation}
\kappa_{\mathrm{L}}^{\mu\nu} = \frac{1}{N\Omega}\sum_{q}C_{q}v_{q}^{\mu}v_{q}^{\nu}\tau_{q},
\label{eq:kappa}
\end{equation}
where $\Omega$ is the unit-cell volume and $\tau_{q} = (2\Gamma_{q}(\omega_{q}))^{-1}$ is the phonon lifetime.
Although RTA can underestimate thermal conductivity of high-$\kappa_{\mathrm{L}}$ materials because it incorrectly considers normal phonon-phonon processes as resistive~\cite{Ward2009}, it has been reported that, in most of the materials, the RTA gives $\kappa_{\mathrm{L}}$ that doesn't differ much from the full solution to the BTE  especially for low-$\kappa_{\mathrm{L}}$ systems~\cite{Lindsay2013,Chernatynskiy2011}.

\begin{figure}[t]
\centering
\includegraphics[width=8.5cm,clip]{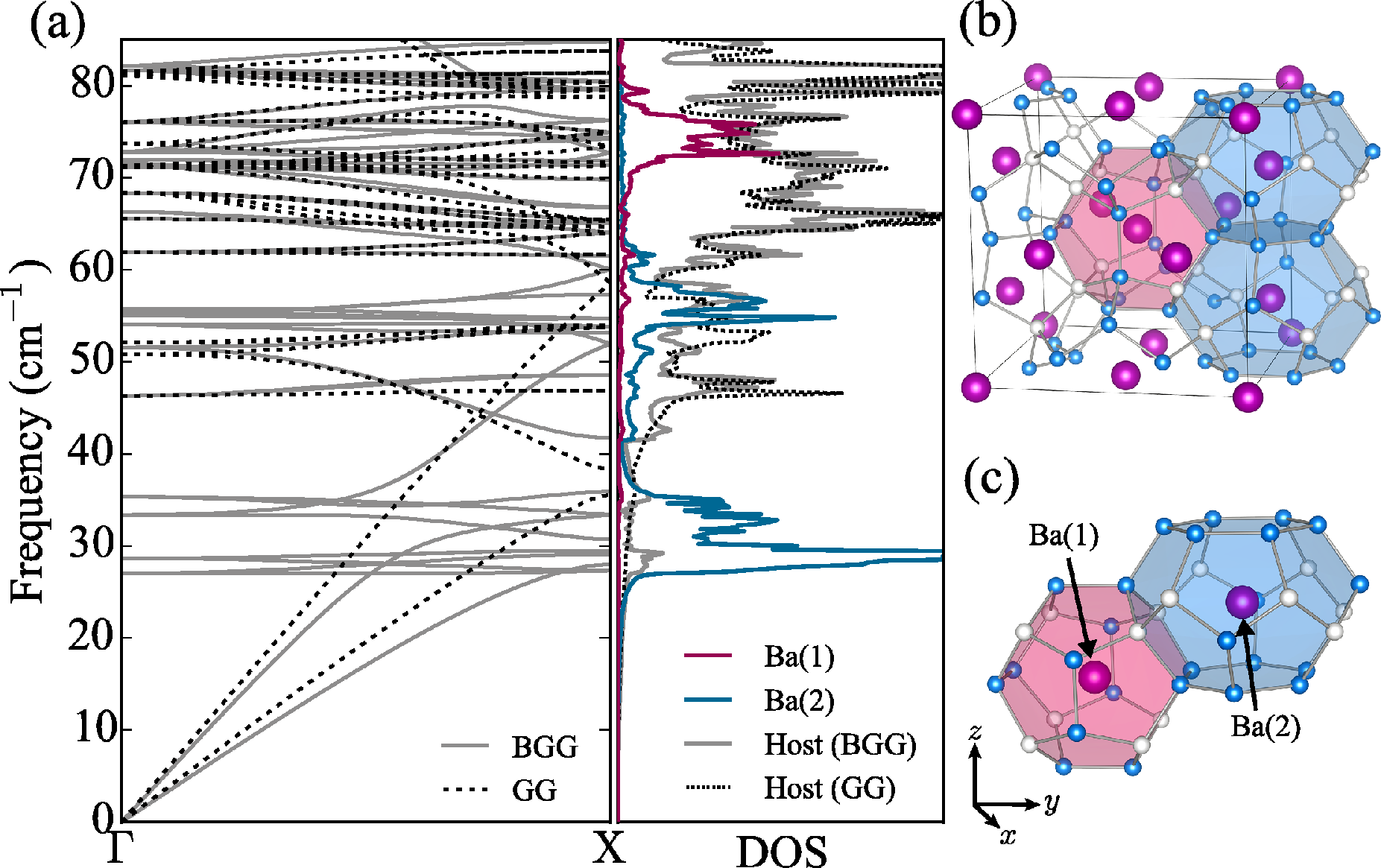}
\caption{(a) Calculated phonon dispersion and projected phonon DOS of BGG (solid) and GG (dotted). 
(b) Crystal structure of BGG. Ba, Ga, and Ge atoms are represented by magenta, white, and blue spheres, respectively. 
(c) Representation of two inequivalent Ba rattlers in dodecahedral (red) and tetrakaidecahedral (blue) polyhedra. }
\label{fig:dispersion}
\end{figure}

The calculation of the phonon linewidth requires $\Phi_{\mu_{1}\mu_{2}}(\ell_{1}\alpha_{1};\ell_{2}\alpha_{2})$ and 
$\Phi_{\mu_{1}\mu_{2}\mu_{3}}(\ell_{1}\alpha_{1};\ell_{2}\alpha_{2};\ell_{3}\alpha_{3})$, which are harmonic and cubic force constants in real space. 
We computed these by the finite-displacement approach \cite{Tadano2014,Esfarjani2008} where atomic forces for displaced configurations were calculated with the VASP code \cite{Kresse1996}, 
with the Perdew-Burke-Ernzerhof (PBE) functional \cite{PBE1996} and the PAW method~\cite{PAW1994,Kresse1999} (see Supplemental Material~\cite{supplement}).
Calculations were performed for a unit cell containing 54 atoms with the crystal symmetry of $Pm\bar{3}n$ (see Figs.~\ref{fig:dispersion}(b) and \ref{fig:dispersion}(c)).
The optimized lattice constant was 10.95~{\AA}, which slightly overestimates the experimental value, 10.76~{\AA}~\cite{Christensen2006}.



The calculated phonon dispersion and the phonon DOS of BGG are shown in Fig.~\ref{fig:dispersion}(a) with solid lines. 
The avoided crossing of the longitudinal acoustic (LA) mode and the rattling modes of Ba(2) atoms inside tetrakaidecahedral cages is clearly observed in the band structure around $\omega_{q} = $ 28-34 cm$^{-1}$. 
The rattling modes in this energy range correspond to the vibration in the $xy$ plane, 
whereas those observed in $\sim$55 cm$^{-1}$ are the vibration of Ba(2) atoms along the $z$ direction.
Experimentally measured Raman shifts of these rattling modes are 31-33 cm$^{-1}$ and 64 cm$^{-1}$ respectively~\cite{Takasu2006},
in good agreement with the computational results. We found that the small underestimation of the vibration energy along direction $z$ can be cured by employing a smaller lattice constant comparable with the experimental value. 
In this study, we neglect the temperature dependence of the Raman spectra reported by Takasu \textit{et al.}~\cite{Takasu2006} due to computational limitations.
We consider such effect is not essential to understand the low $\kappa_{\mathrm{L}}$ of BGG, but could be important to explain the temperature dependence of $\kappa_{\mathrm{L}}$ especially for off-center systems.
To better understand the localization of the Ba motions and their hybridization with Ga/Ge atoms, 
we estimated the participation ratio $P_{q} = (\sum_{\alpha} |\bm{u}_{\alpha}(q)|^{2})^{2}/N_{\alpha}\sum_{\alpha}|\bm{u}_{\alpha}(q)|^{4}$ where $\bm{u}_{\alpha}(q) = M_{\alpha}^{-1/2}\bm{e}_{\alpha}(q)$ as in Ref.~\onlinecite{Pailhes:2014fe}.
We then obtained small $P_{q}$ ($\sim 0.1$) for all the rattling modes including vibrations of Ba(1) atoms inside dodecahedral cages observed in 70-80 cm$^{-1}$, thus indicating the localized nature of vibrations of Ba guests (see Supplemental Material~\cite{supplement}).
Localization of guest atoms is significant in the rattling modes around $\omega_{q} =$ 28-34 cm$^{-1}$, whereas a little hybridization with cage atoms is observed in the higher energy rattling modes, in accordance with previous computational results~\cite{Euchner:2012er,Pailhes:2014fe}.


To investigate the effect of Ba rattlers directly, we also calculated phonon properties of an empty clathrate Ga$_{16}$Ge$_{30}$(GG).
Since Ba atoms are necessary as electron donors to complete the $sp^{3}$ framework of host cages, 
one cannot perform first-principles calculations directly on the fictitious material GG. 
Therefore, we estimated harmonic and cubic force constants of GG 
from the displacement-force data set obtained for the original system BGG by neglecting force constants related to Ba guests. 
We found that this procedure can extract force constants of the hypothetical system GG 
without degrading the numerical accuracy of host-host force constants (see Supplemental Material~\cite{supplement}).
By using the extracted force constants of GG, we can turn off host-guest interactions and the crossing behavior is recovered as shown in Fig.~\ref{fig:dispersion}(a) (dotted lines).


\begin{figure}[t]
\centering
\includegraphics[width=8.0cm,clip]{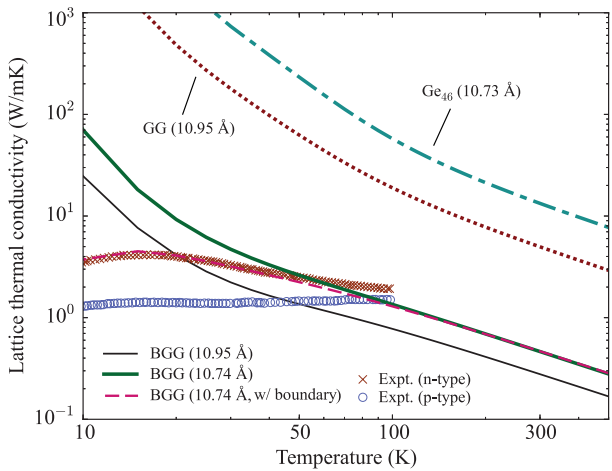}
\caption{Temperature dependence of the lattice thermal conductivity calculated for filled (BGG) and empty (GG, Ge$_{46}$) clathrates. For BGG, we also show the result with the boundary effect considered by the Matthiessen's rule with the grain size of $L = 2~\mu$m, which is comparable with experimental results (symbols)~\cite{Avila2006}.}
\label{fig:kl_all}
\end{figure}

Figure ~\ref{fig:kl_all} shows the temperature dependence of $\kappa_{\mathrm{L}}$ calculated by the BTE-RTA [Eq.~(\ref{eq:kappa})]. 
The lattice thermal conductivity of BGG calculated with 6$\times$6$\times$6 $\bm{q}$ points was 0.78 W/mK at 100 K~\cite{note_kl_exp}. The value may slightly be increased by using denser $\bm{q}$ grids, 
although such calculations were not performed due to computational limitations.
In addition, we observed that $\kappa_{\mathrm{L}}$ of BGG is sensitive to the lattice constant, 
and a smaller lattice constant corresponding to the experimental value yielded $\kappa_{\mathrm{L}} = 1.35$ W/mK.
This value agrees well with experimental values 1.5 W/mK (p-type) and 1.9 W/mK (n-type) \cite{Avila2006},
thus indicating the validity of the BTE-RTA with considering dominant three-phonon scattering processes even for the complex system BGG.
The thermal conductivity of a empty system GG calculated with 8$\times$8$\times$8 $\bm{q}$ points was 18.94 W/mK at 100 K (16.74 W/mK with 6$\times$6$\times$6 $\bm{q}$ points), which is more than 20 times greater than that of BGG. 
We also estimated $\kappa_{\mathrm{L}}$ of another empty clathrate Ge$_{46}$ and obtained a higher value 57.59 W/mK (8$\times$8$\times$8 $\bm{q}$ points), which can be attributed to the smaller anharmonicity of the Ge framework.
These results clearly show the impact of rattlers on $\kappa_{\mathrm{L}}$ of type-I clathrates, 
which was also demonstrated by previous classical molecular dynamics studies on a hypothetical material Sr$_{6}$Ge$_{46}$~\cite{Dong2001}. 
In Fig.~\ref{fig:kl_all}, we also show a theoretical result of BGG with the effect of phonon-boundary scattering considered by the Matthiessen's rule, for which the phonon lifetime in Eq.~(\ref{eq:kappa}) is substituted by $\tau_{q,\mathrm{eff}}^{-1}=\tau_{q}^{-1} + 2v_{q}/L$ with the grain size of $L = 2~\mu$m. 
The result agrees well with the experimental values of the n-type sample~\cite{Avila2006} in a wide temperature range. The lower thermal conductivity of the p-type sample can be attributed to the off-center rattling motion as suggested by Fujiwara and co-authors~\cite{PhysRevB.85.144305}.


\begin{figure}[t]
\centering
\includegraphics[width=8.5cm,clip]{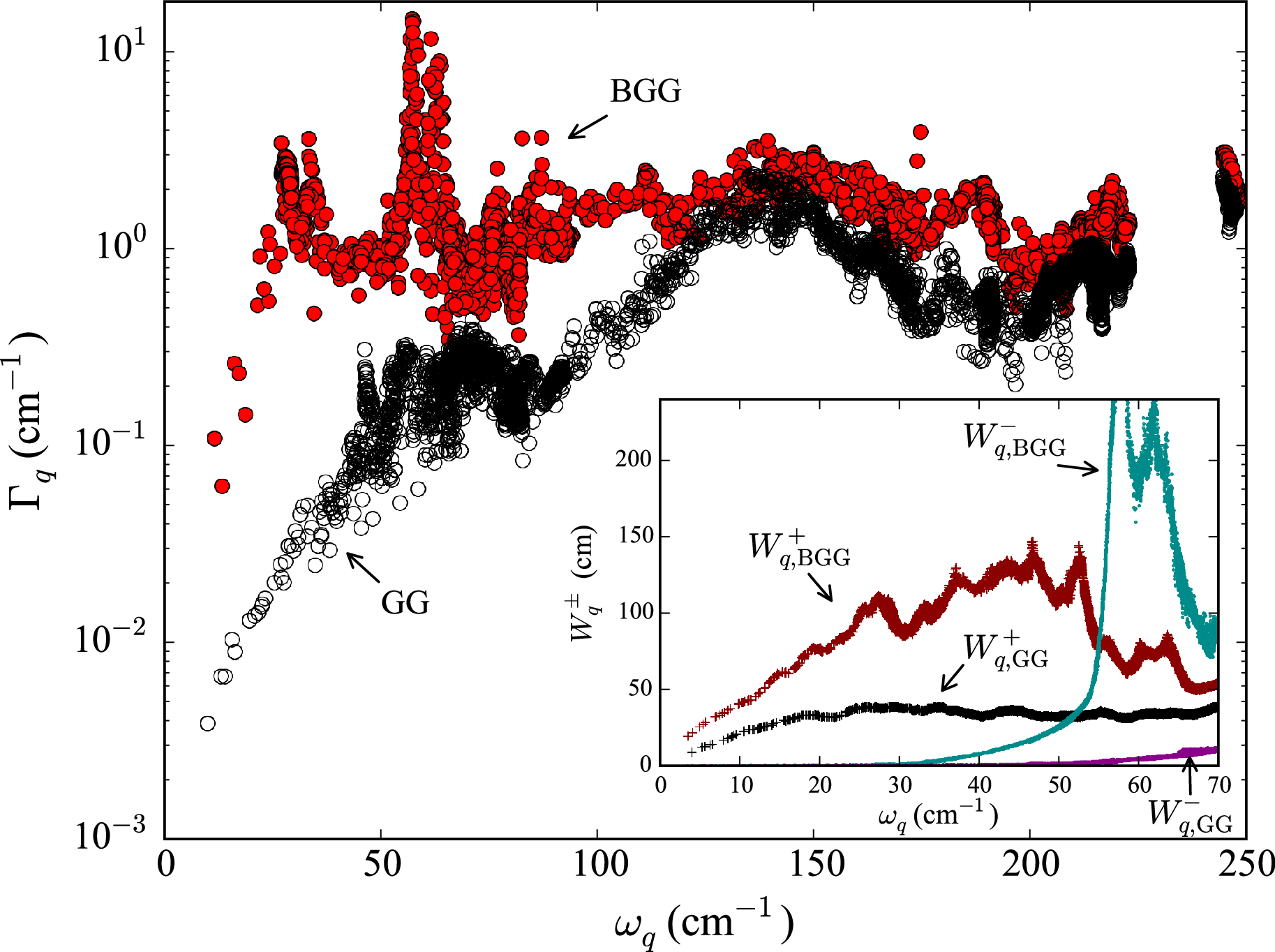}
\caption{Calculated phonon linewidth of BGG (filled circle) and GG (open circle) at 300 K. 
Inset: Energy- and momentum-conserving phase space $W_{q}^{\pm}$ in the low energy region at 300 K.}
\label{fig:linewidth}
\end{figure}

\begin{figure}[t]
\centering
\includegraphics[width=8.5cm,clip]{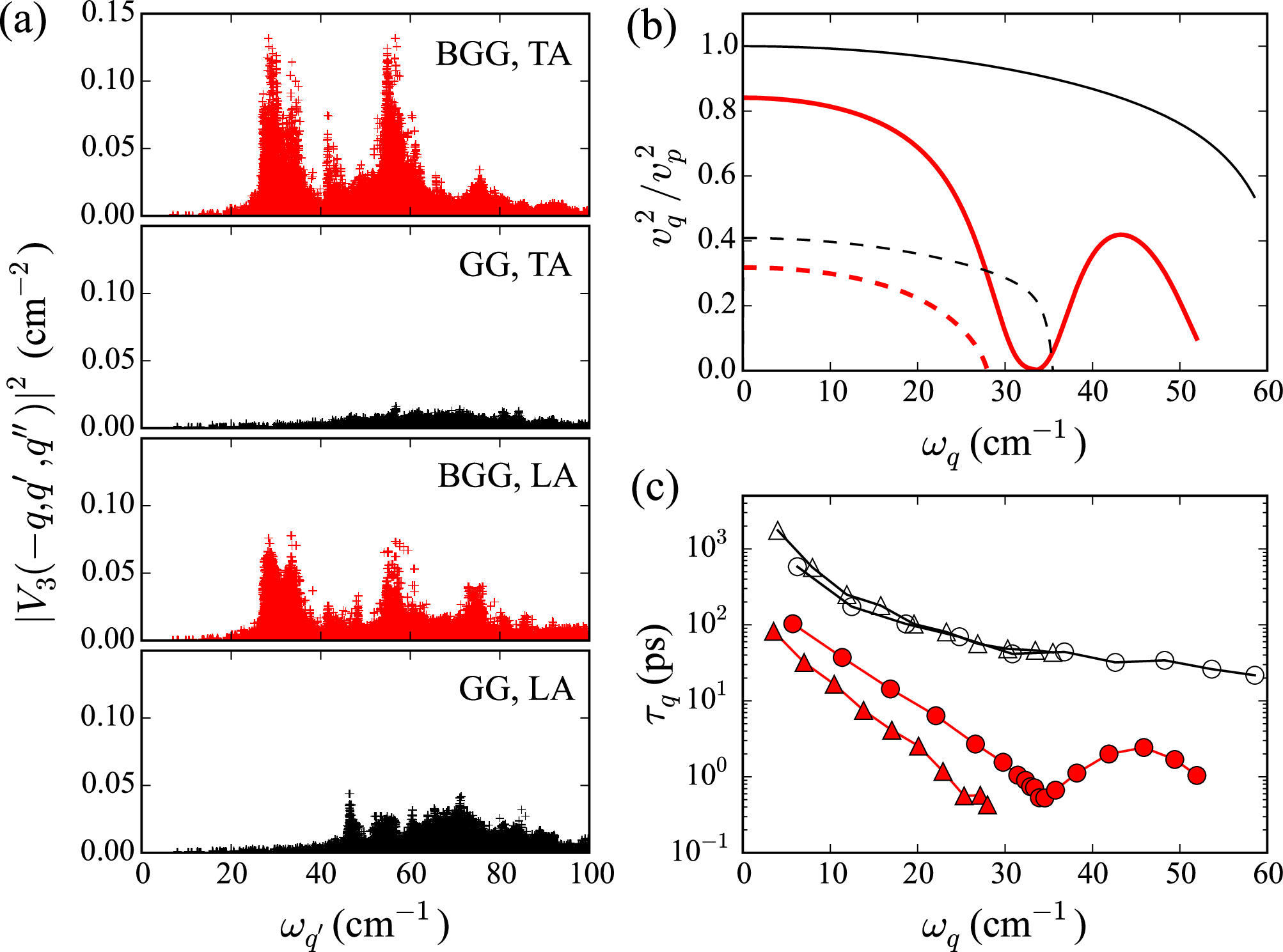}
\caption{Calculated phonon properties of acoustic and the anti-crossing rattling modes along $[100]$ direction.
(a) Magnitude of anharmonic interactions of the TA and LA modes at $\bm{q} = (0.1, 0, 0)$.
(b) Square group velocity of transverse (dashed) and longitudinal (solid) modes of BGG (thick) and GG (thin) 
normalized by the square sound velocity $v_{p}^{2}$. 
(c) Phonon lifetime of transverse (triangle) and longitudinal (circle) modes of BGG (filled symbol) and GG (empty symbol) at 300 K.}
\label{fig:resonant}
\end{figure}

To understand the microscopic origin of the 20-fold reduction in $\kappa_{\mathrm{L}}$, we analyzed the role of rattlers on the phonon lifetime $\tau$.
In Fig.~\ref{fig:linewidth}, we compare the phonon linewidth $\Gamma_{q}$ of BGG and GG at room temperature calculated by Eq.~(\ref{eq:Gamma}), 
which is inversely proportional to the phonon lifetime; $\tau_{q} = (2\Gamma_{q})^{-1}$. 
Contrary to the resonant scattering picture where one expects $\Gamma_{q}(\omega) \propto \frac{\omega^{2}}{(\omega^{2}-\omega_{0}^{2})^{2}}$ with a localized rattling frequency $\omega_{0}$~\cite{Pohl1962},
the phonon linewidth of BGG is increased from that of GG in the entire frequency range.
Therefore, phonon scattering by rattlers is \textit{not} resonant.
The difference is significant in the low-frequency region ($< 120$ cm$^{-1}$), and the 10-fold increase in $\Gamma_{q}$ is observed even for low-frequency acoustic phonons.
To elucidate the reason of this considerable increase in $\Gamma_{q}$ due to rattlers, 
we also estimated the energy- and momentum-conserving phase space for three-phonon scattering processes
\begin{equation}
W^{\pm}_{q} = {\sum_{q',q''}}^{\prime} 
\left\{
    \begin{array}{c}
      n_{q''} - n_{q'} \\
      n_{q'} + n_{q''} + 1
    \end{array}
  \right\}
\delta(\omega_{q}-\omega_{q'}\pm \omega_{q''}).
\end{equation}
Here, $W_{q}^{+}$ and $W_{q}^{-}$ are the phase space corresponding to absorption and emission of phonon $q$, respectively,
and the summation is restricted to the pairs $(q',q'')$ satisfying $\bm{q}+\bm{q}'+\bm{q}''=n\bm{G}$.
At room temperature, the phase space of absorption $W_{q}^{+}$ is significantly greater than that of emission $W_{q}^{-}$ 
for low-energy modes as shown in Fig.~\ref{fig:linewidth} inset. 
Moreover, additional phonon modes induced by rattlers increase the phase space $W_{q}^{+}$ for low-energy acoustic phonons; 
$W_{q,\mathrm{BGG}}^{+}$ is greater than $W_{q,\mathrm{GG}}^{+}$ by more than factor 2. 
However, this increase in $W^{\pm}_{q}$ is not sufficient to explain the 10-fold increase in $\Gamma_{q}$.
Actually, the increase in the matrix element $|V_{3}(-q,q',q'')|^{2}$ appearing in Eq.~(\ref{eq:Gamma}) is also significant in BGG.
In Fig.~\ref{fig:resonant}(a), we show the difference in the matrix element $|V_{3}(-q,q',q'')|^{2}$ of heat-carrying TA (transverse acoustic) and LA (longitudinal acoustic) modes at $\bm{q} = (0.1, 0, 0)$ whose frequencies are 8.0 cm$^{-1}$ and 11.4 cm$^{-1}$, respectively. 
As clearly seen in the figure, the matrix elements of BGG are increased at the frequencies $\omega_{q'}$ corresponding to the vibrational energies of localized rattling modes (see Fig.~\ref{fig:dispersion}).
This indicates the significant role of rattlers for enhancing $|V_{3}(-q,q',q'')|^{2}$, a measure of anharmonicity of the system.
Not only the LA mode, but also the TA mode at small $\bm{q}$ is strongly affected by the localized Ba motions, which is made possible by the anharmonic effects.
In the frequency above 100 cm$^{-1}$, we observed no significant difference in $|V_{3}(-q,q',q'')|^{2}$ between BGG and GG. 
Therefore, the difference of the matrix elements in the low-frequency region should be essential for 
understanding the 10-fold change in $\tau$ and very low-$\kappa_{\mathrm{L}}$ of BGG.
It should be noted that our results shown in Figs.~\ref{fig:linewidth} and \ref{fig:resonant}
clearly support the increased Umklapp scattering scenario proposed by Lee \textit{et al.}~\cite{HLee:2006ej},
and hereby question the validity of the phononic filter mechanism recently proposed by Euchner and Pailh\`{e}s~\cite{Euchner:2012er,Pailhes:2014fe}.


Next, we discuss the effect of rattlers on the group velocity $v$ of heat-carrying acoustic modes. 
To this end, we calculated group velocities and phonon lifetimes of the TA and LA modes (and the anti-crossing rattling mode for BGG) along $[100]$ direction.
In Fig.~\ref{fig:resonant}(b), we compare the square group velocities of the transverse and longitudinal modes of BGG and GG.
The values are normalized by the group velocity of the LA mode of GG, $v_{p} = 4100$ m/s.
It can be seen that the group velocity of the LA mode is significantly reduced around $\omega_{q} = 33$ cm$^{-1}$ due to
the flattening of the band dispersion, whereas the TA mode is less affected.
The reduction in $v$ seems great, but is too local in energy to explain the 20-fold reduction of the
integrated quantity $\kappa_{\mathrm{L}} \propto \int d\omega D(\omega) C(\omega) v^{2}(\omega) \tau(\omega)$, where $D(\omega)$ is the phonon DOS.
From the $\omega$-integral of $v^{2}_{\mathrm{LA}}(\omega)$, we can roughly estimate that the 
reduction in $v$ due to rattlers results in a reduction in $\kappa_{\mathrm{L}}$ by no more than factor 2.
Moreover, if the frequency dependence of $\tau(\omega)$ is considered, 
the significance of the group velocity reduction becomes much less.
In Fig.~\ref{fig:resonant}(c), we also compare the phonon lifetimes of the acoustic modes at room temperature. 
As already noted in Fig.~\ref{fig:linewidth}, the rattler-induced reduction in $\tau(\omega)$ occurs in the wide frequency range 
rather than in a limited range around 33 cm$^{-1}$.
This is because the phonon-rattler scattering is described by precise three-phonon processes rather than the phenomenological resonant scattering. 
The calculated phonon lifetimes of the TA and LA modes at room temperature are $\tau \sim 1$ ps near the avoided-crossing point and increase rapidly as decreasing $\bm{q}$, in good agreements with previous INS studies~\cite{Euchner:2012er,Pailhes:2014fe}.
We also estimated the mean-free-path $\ell_{q} = |v_{q}| \tau_{q}$ of acoustic phonons, 
and found $\ell_{q}$ of low-energy modes can be as long as 100 nm even at room temperature. 
In the previous INS study~\cite{Christensen2008}, phonon mean-free-paths in BGG were assumed to be restricted by the separation of rattlers ($\ell_{q} \sim 5.5$ \AA). Such an assumption, however, is artificial because we observed $\ell_{q} > 1$ nm in a wide range of frequency.


To summarize, we performed anharmonic lattice dynamics calculations of a type-I clathrate BGG from first principles and
analyzed the effect of on-center rattlers on the relaxation time $\tau$ and the group velocity $v$ of acoustic phonons.
Then, we observed massive reduction of $\tau$ in a wide frequency range, thus indicating that the resonant scattering scenario is not appropriate to describe phonon scatterings in BGG.
This 10-fold reduction in $\tau$ was achieved due to the enhancement of the anharmonicity and the energy- and momentum-conserving phase space induced by rattlers. 
The nonresonant and significant reduction of $\tau$ is crucial to understand unusually low lattice thermal conductivities ($\kappa_{\mathrm{L}}\sim 1$ W/mK) of clathrates, 
whereas the effect of the group velocity is secondary because its reduction occurs only in a limited frequency range.


This study is supported by a Grant-in-Aid for Scientific Research on Innovative Areas 
``Materials Design through Computics: Complex Correlation and Non-Equilibrium Dynamics'' 
from the Ministry of Education, Culture, Sports, Science and Technology (MEXT) of Japan, 
and is partially supported by Tokodai Institute for Element Strategy (TIES). 
The computation in this work has been done using the facilities of the Supercomputer Center, 
Institute for Solid State Physics, The University of Tokyo.

\bibliography{paper2}

\begin{thebibliography}{34}
\bibitem [{not()}]{note_kl_exp}%
  \BibitemOpen
  \href@noop {} {}\bibinfo {note} {Here we compare the experimental and
  calculated $\kappa_{\mathrm{L}}$ at 100 K where phonon-phonon scatterings
  dominate the phonon scattering probability and the bipolar contribution,
  which is noticeable in experimental $\kappa_{\mathrm{L}}$ at higher
  temperature, is negligible.}\BibitemShut {Stop}%
\end{thebibliography}

\widetext
\clearpage

\setcounter{equation}{0}
\setcounter{figure}{0}
\setcounter{table}{0}
\makeatletter
\renewcommand{\theequation}{S\arabic{equation}}
\renewcommand{\thefigure}{S\arabic{figure}}
\renewcommand{\bibnumfmt}[1]{[S#1]}
\renewcommand{\citenumfont}[1]{S#1}

\begin{center}
\textbf{\large Supplemental Material}
\end{center}

\subsection{A. Computational details}

In this work, {\it ab initio} DFT calculations were conducted using the VASP code~\cite{Kresse1996S},
which uses the projector augmented wave (PAW) method~\cite{PAW1994S,Kresse1999S}.  The adapted PAW
potentials treat the Ba $5s^{2}5p^{6}6s^{2}$, Ga $3d^{10}4s^{2}4p^{1}$, and Ge $3d^{10}4s^{2}4p^{2}$
shells as valence states, and the cutoff energy of 320 eV was employed. For the exchange-correlation
functional, we employed the Perdew-Burke-Ernzerhof (PBE) functional~\cite{PBE1996S}. The Brillouin-zone 
integration was performed with the 4$\times$4$\times$4 Monkhorst-Pack $k$
grid~\cite{MonkhorstPack}, with which the smearing scheme proposed by Methfessel and Paxton was used
with width 0.2 eV~\cite{MethfesselPaxton}. We optimized the lattice constant of a unit cell
containing 54 atoms with the crystal symmetry of $Pm\bar{3}n$ and obtained $a = 10.95$~{\AA}, which
slightly overestimates the experimental value, 10.76~{\AA}~\cite{Christensen2006S}.  The internal
coordinates were also optimized so that the force convergence threshold of 0.1 meV/{\AA} was
satisfied.

The harmonic and cubic force constants were calculated by the finite-displacement
approach~\cite{Esfarjani2008S,Tadano2014S}.  To calculate harmonic terms, we displaced an atom from
its equilibrium position by 0.02~{\AA} and calculated Hellmann-Feynman forces.  For cubic terms, two
atoms were simultaneously displaced along various directions by 0.04~\AA.  We introduced a cutoff
radius for the cubic interactions so that Ba-cage (Ga/Ge) and cage-cage interactions up to  the
nearest-neighbor shells were considered. The number of symmetrically inequivalent force constants
were found to be 320 and 384 for harmonic and cubic terms, respectively, which were estimated by
least-square fitting. For the fitting procedure and the subsequent phonon calculations, we employed
the ALAMODE package~\cite{Tadano2014S,alamode}, an open source software developed for anharmonic
phonon calculations. The phonon calculations were conducted with the gamma-centered
20$\times$20$\times$20 $q$ grid except for  phonon linewidth and thermal conductivity, for which we
employed 6$\times$6$\times$6 and 8$\times$8$\times$8 $q$ grids for filled and empty clathrates,
respectively. The delta function appearing in DOS-related quantities was evaluated by the
tetrahedron method~\cite{PhysRevB.49.16223}.

\clearpage

\subsection{B. Force constants of the empty clathrate Ga$_{16}$Ge$_{30}$}
\label{sec:fcs}

\begin{figure}[h]
\centering
\subfigure[Harmonic force constants]{
 \includegraphics[width=0.48\textwidth,clip]{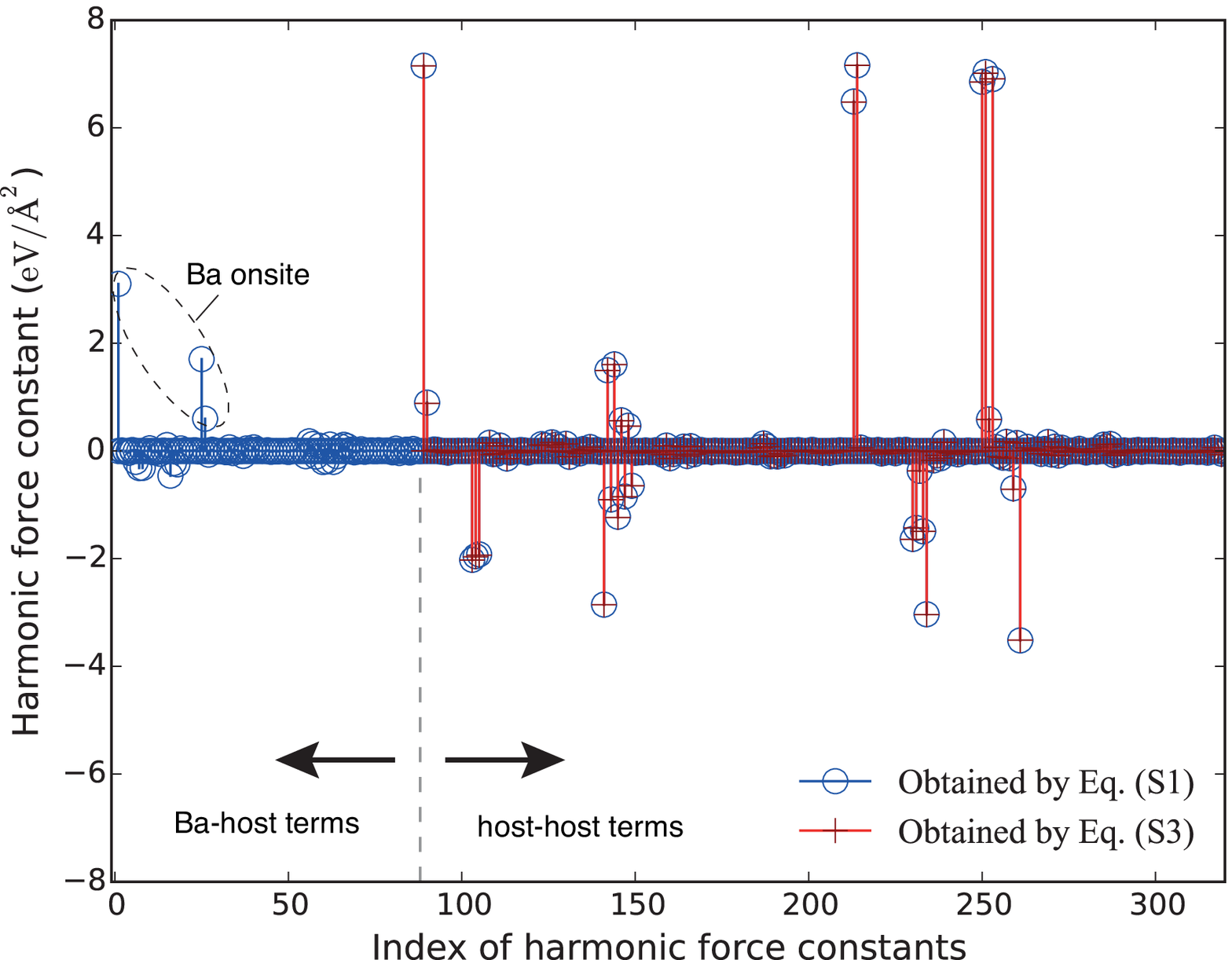}   
 \label{fig:fc2_compare}
}
\hfill
\subfigure[Cubic force constants]{
 \includegraphics[width=0.48\textwidth,clip]{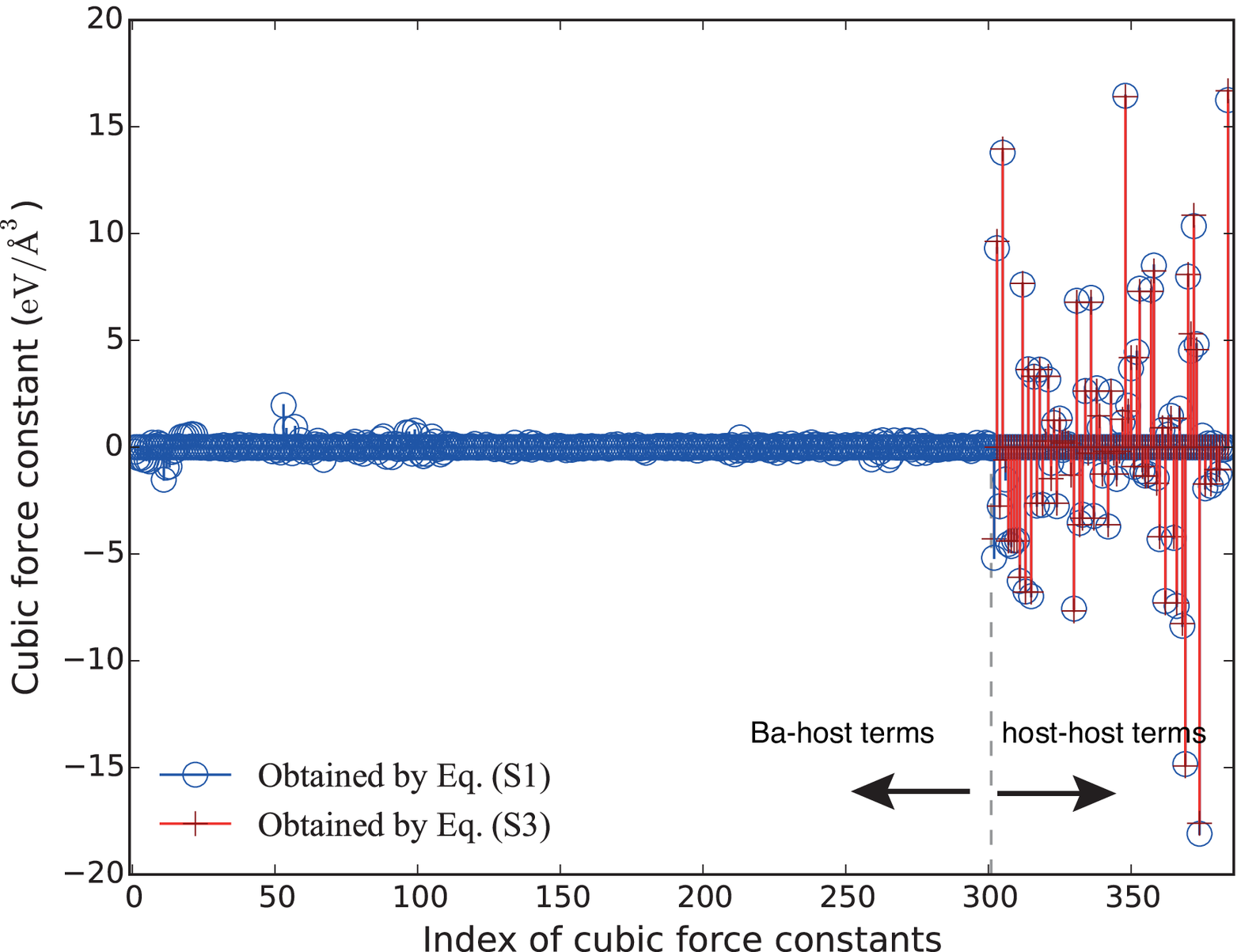}
 \label{fig:fc3_compare}
}
\caption{Comparison of the force constants calculated by Eq.~(\ref{eq:fitting}) and Eq.~(\ref{eq:fitting2}).}
\label{fig:fcs_compare}
\end{figure}

We estimated harmonic and cubic force constants of an empty clathrate Ga$_{16}$Ge$_{30}$ (GG) from the displacement-force
data set obtained by DFT calculations of the filled clathrate Ba$_{8}$Ga$_{16}$Ge$_{30}$ (BGG). 
In the case of BGG, the force constants were extracted by solving the following least square problem:
\begin{equation}
\text{Minimize} \ \ \ \chi^{2} = \sum_{m}\sum_{i} \| F_{i,m}^{\mathrm{DFT}} - F_{i,m}^{\mathrm{Model}} \|^{2},
\label{eq:fitting}
\end{equation}
where 
\begin{equation}
F_{i,m}^{\mathrm{Model}} = -\sum_{j}\Phi_{i,j}u_{j} - \frac{1}{2}\sum_{j,k}\Phi_{i,j,k}u_{j}u_{k}.
\end{equation}
Here, $i$, $j$, and $k$ are the triplet index of $\{\ell,\kappa,\mu\}$, and $m$ in Eq.~(\ref{eq:fitting}) is 
the index of reference data set prepared by DFT calculations. To obtain force constants of GG, 
harmonic terms $\Phi_{i,j}$ and cubic terms $\Phi_{i,j,k}$ involving Ba rattlers were treated as zero, 
and the remaining parameters were determined by solving the following problem:
\begin{equation}
\text{Minimize} \ \ \ \chi^{2} = {\sum_{m}}^{\prime}\sum_{i\in\mathrm{Ga,Ge}} \| F_{i,m}^{\mathrm{DFT}} - F_{i,m}^{\mathrm{Model}} \|^{2}.
\label{eq:fitting2}
\end{equation}
Here, the summation over the reference data is restricted to the configuration where only host atoms are displaced. 
Following this procedure, we were able to turn off the host-guest interactions without sacrificing the accuracy of 
host-host force constants as shown in Fig.~\ref{fig:fcs_compare}.

\clearpage

\subsection{C. Phonon dispersion}

\begin{figure}[!h]
\centering
\includegraphics[width=\textwidth,clip]{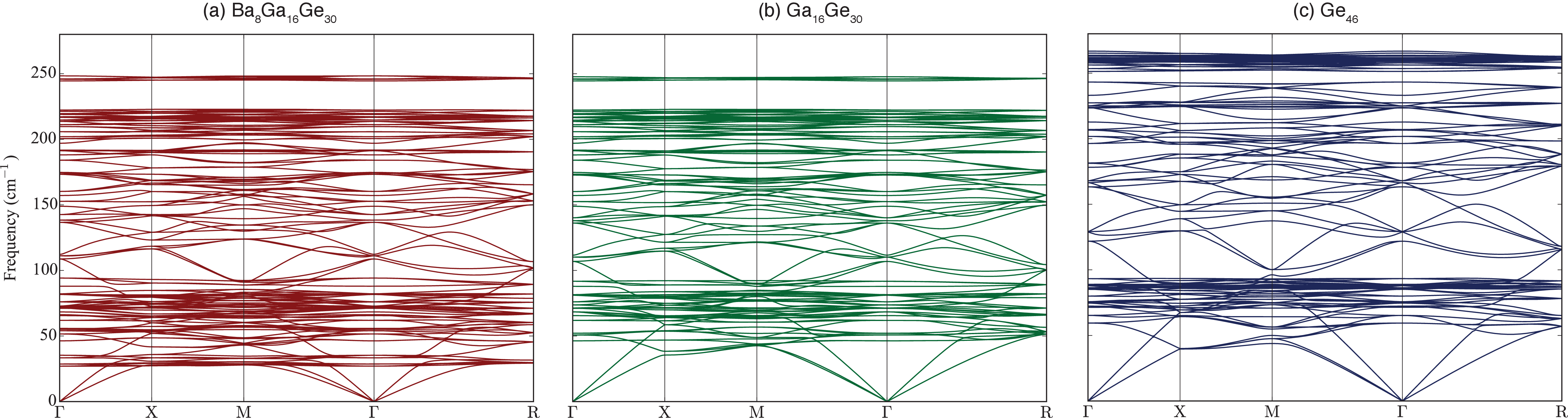}
\caption{Phonon dispersion of (a) Ba$_{8}$Ga$_{16}$Ge$_{30}$, (b) Ga$_{16}$Ge$_{30}$, 
and (c) Ge$_{46}$ along the high-symmetry lines of the Brillouin zone.}
\label{fig:dispersion_all}
\end{figure}

\begin{figure}[!h]
\centering
\includegraphics[width=\textwidth,clip]{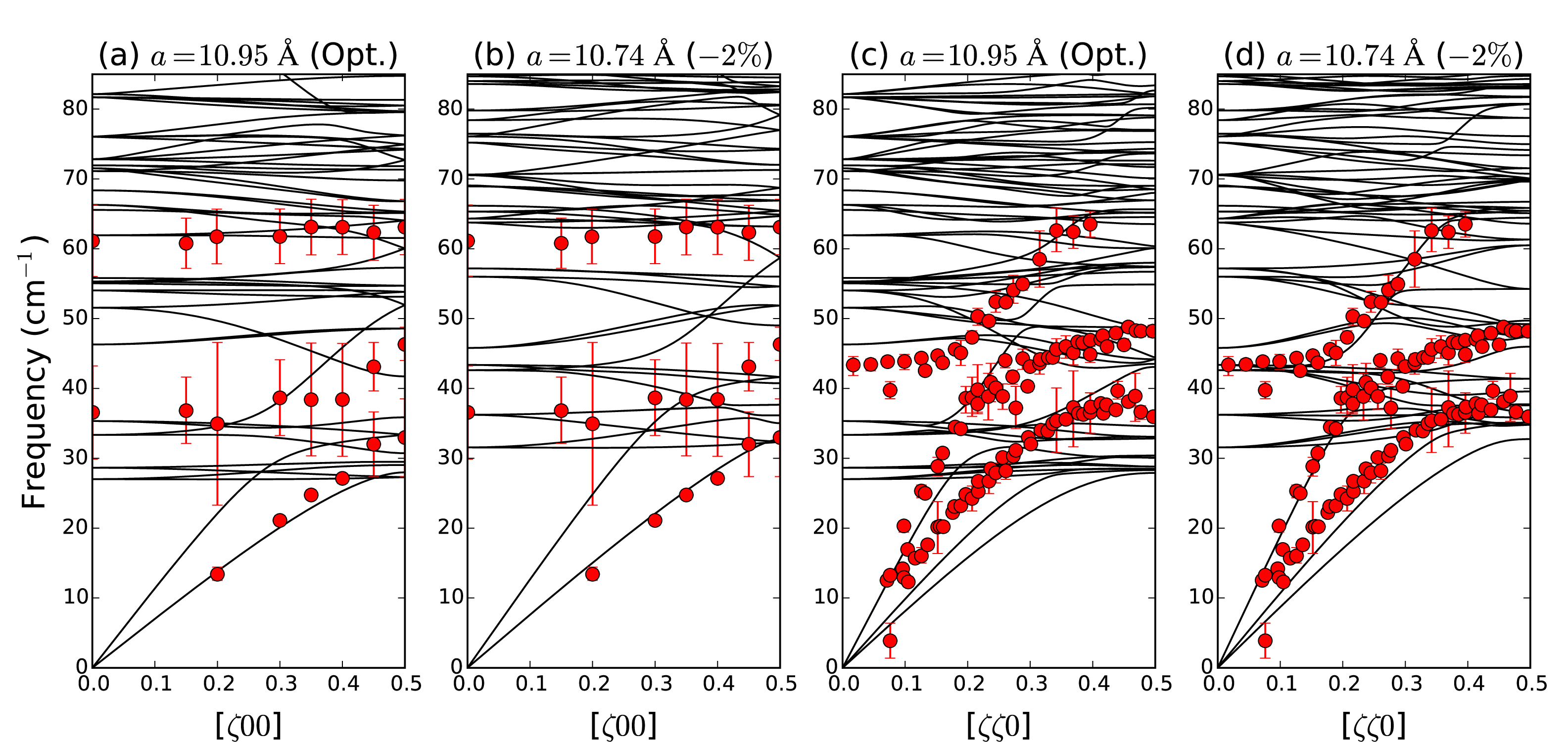}
\caption{Calculated phonon dispersion of Ba$_{8}$Ga$_{16}$Ge$_{30}$ compared with experimental results (red circle) 
obtained from Refs. \onlinecite{Christensen2008,Lee2007}. Agreement is satisfactory especially when we employ
the lattice constant comparable with the experimental value, 10.76~\AA.}
\label{fig:dispersion_exp_compare}
\end{figure}

Figure~\ref{fig:dispersion_all} shows the calculated phonon dispersion relations of filled and empty
clathrates along the high-symmetry lines of the Brillouin zone.  Using the method described in
Sec. B, low-frequency Ba modes are removed in Ga$_{16}$Ge$_{30}$.  We also calculated
the phonon dispersion relation of Ge$_{46}$ for comparison employing the same computational
conditions.  Phonon frequencies of Ge$_{46}$ are higher than those of Ge$_{16}$Ge$_{30}$, which can
be attributed to the smaller lattice constant of Ge$_{46}$ (10.73 \AA).  In this study, we compared
phonon properties of Ba$_{8}$Ga$_{16}$Ge$_{30}$ and Ga$_{16}$Ge$_{30}$  because it can highlight the
role of intercalated Ba atoms more clearly than comparing Ba$_{8}$Ga$_{16}$Ge$_{30}$ and Ge$_{46}$.
In Fig.~\ref{fig:dispersion_exp_compare}, we compare experimental and theoretical phonon dispersion
of Ba$_{8}$Ga$_{16}$Ge$_{30}$ along $[100]$ and $[110]$ directions.


\clearpage

\subsection{D. Participation ratio}

\begin{figure}[!h]
\begin{minipage}{0.49\hsize}
\centering
\includegraphics[width=0.95\textwidth,clip]{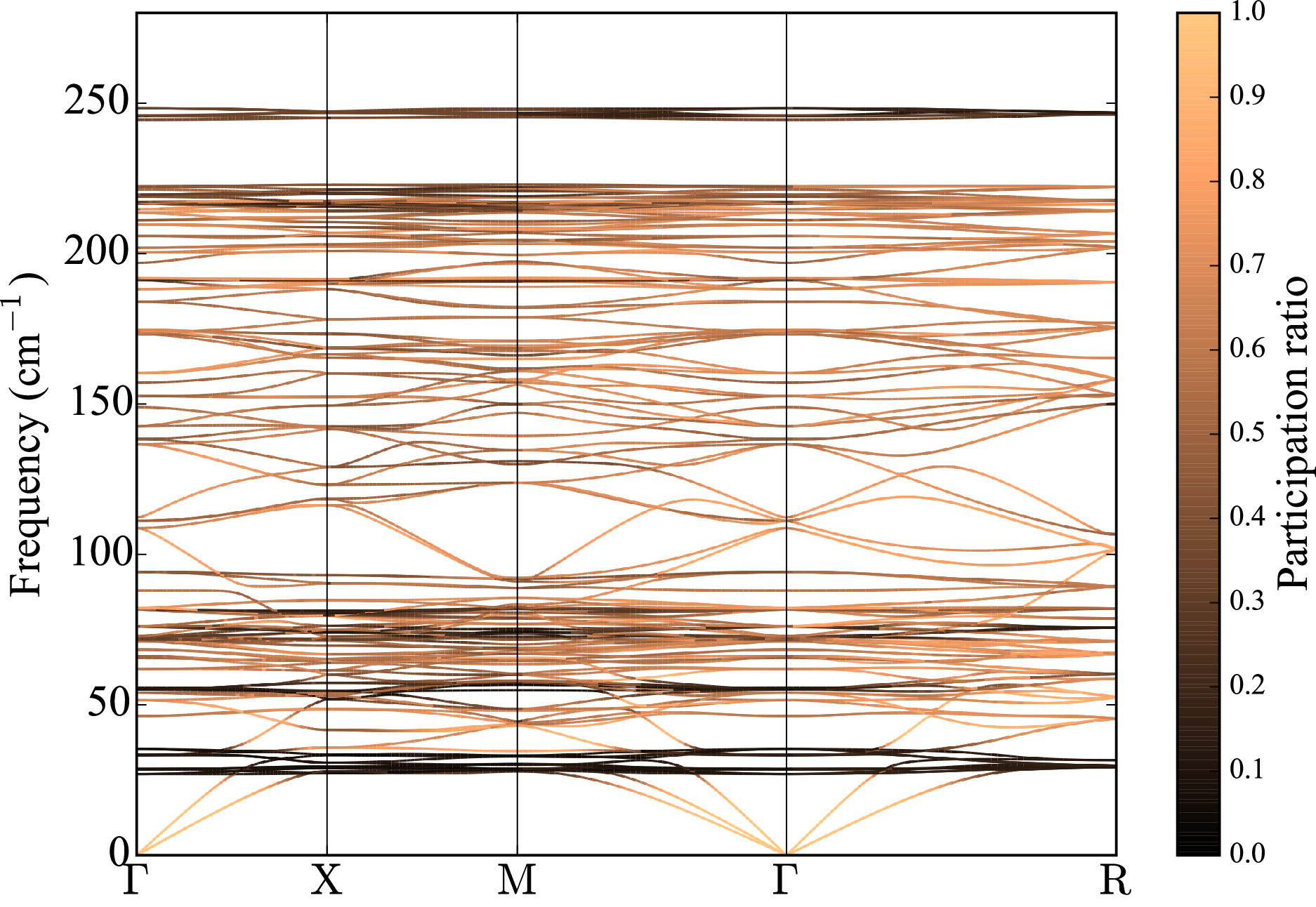}
\caption{Participation ratio of Ba$_{8}$Ga$_{16}$Ge$_{30}$ shown as a color map.}
\label{fig:participation_ratio}
\end{minipage}
\begin{minipage}{0.49\hsize}
\centering
\includegraphics[width=0.9\textwidth,clip]{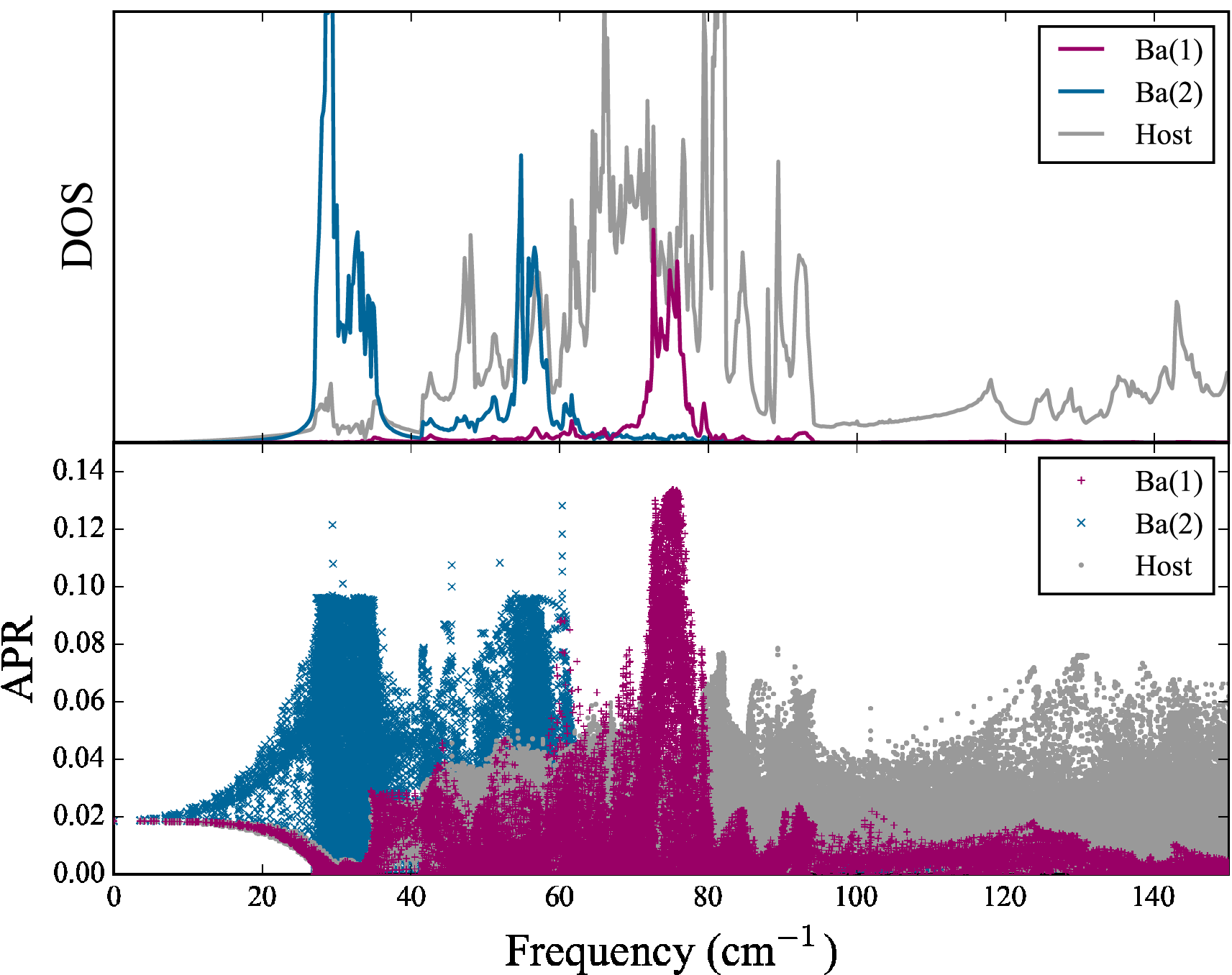}
\caption{Atomic participation ratio (APR) of Ba$_{8}$Ga$_{16}$Ge$_{30}$ (bottom) shown with the atom-projected phonon DOS (top).}
\label{fig:atomic_participation_ratio}
\end{minipage}
\end{figure}

To analyze the localization of Ba atoms and their hybridization with host atoms, we calculated the
participation ratio~\cite{Euchner2012S,Pailhes2014S} 
\begin{equation} P_{q} =
\left(\sum_{\alpha}^{N_{\alpha}} \frac{|\bm{e}_{\alpha}(q)|^{2}}{M_{\alpha}}\right)^{2} \Bigg/
N_{\alpha} \sum_{\alpha}^{N_{\alpha}} \frac{|\bm{e}_{\alpha}(q)|^{4}}{M_{\alpha}^{2}},
\end{equation} 
where $N_{\alpha}$ is the number of atoms in the primitive cell. $P_{q}$ is close to
1 for extended modes $q$, whereas it is an order of $N_{\alpha}^{-1}$ for localized modes.  As shown
in Fig.~\ref{fig:participation_ratio}, $P_{q}$ is small at $\omega_{q}=$ 28-34 cm$^{-1}$, $\sim$55
cm$^{-1}$, and 70-80 cm$^{-1}$, which correspond to the peak positions in the atom-projected
phonon DOS.  We also estimated the atomic participation ratio (APR)~\cite{Pailhes2014S}
\begin{equation} P_{q,\kappa} =
\frac{|\bm{e}_{\alpha}(q)|^{2}}{M_{\alpha}} \Bigg/ \left( N_{\alpha} \sum_{\alpha}^{N_{\alpha}}
\frac{|\bm{e}_{\alpha}(q)|^{4}}{M_{\alpha}^{2}} \right)^{1/2}, 
\end{equation} which satisfies $(\sum_{\kappa}P_{q,\kappa})^{2} = P_{q}$.
As clearly shown in Fig.~\ref{fig:atomic_participation_ratio}, the phonon modes at $\omega_{q}\sim$ 30 cm$^{-1}$ can be identified as 
localized modes of Ba(2) atoms inside the large cages.
Another localized mode of Ba(2) atoms can be observed around $\omega_{q}\sim$ 55 cm$^{-1}$ with a little hybridization with Ba(1) and host atoms.
Ba(1) atoms in the small cages also yield localized modes around 70-80 cm$^{-1}$.

%

\end{document}